\documentclass[%
 reprint,
superscriptaddress,
 amsmath,amssymb,
 aps,
floatfix,
]{revtex4-2}

\usepackage{graphicx}
\usepackage{dcolumn}
\usepackage{bm}
\usepackage{amsmath,amssymb}
\usepackage{textcomp,mathcomp}

\begin{document}

\preprint{APS/123-QED}

\title{Short-range order and increased transition temperature in LiVO$_2$ with weakened trimer frustration}
\author{K. Kojima}
\affiliation{Department of Applied Physics, Nagoya University, Nagoya 464-8603, Japan}
\author{N. Katayama}\thanks{Corresponding author.}\email{katayama.naoyuki.m5@f.mail.nagoya-u.ac.jp}
\affiliation{Department of Applied Physics, Nagoya University, Nagoya 464-8603, Japan}
\author{Y. Matsuda}
\affiliation{Department of Applied Physics, Nagoya University, Nagoya 464-8603, Japan}
\author{M. Shiomi}
\affiliation{Department of Applied Physics, Nagoya University, Nagoya 464-8603, Japan}
\author{R. Ishii}					
\affiliation{The Institute for Solid State Physics, Tokyo University, Tokyo 277-8581, Japan}
\author{H. Sawa}					
\affiliation{Department of Applied Physics, Nagoya University, Nagoya 464-8603, Japan}
\date{\today}

\begin{abstract}
Vanadium atoms in layered LiVO$_2$ form in-plane periodic vanadium trimers at low temperatures, but the trimers appear randomly in the stacking direction because there are many trimer configurations with comparable lattice energy. We detailed an original modeling scheme to represent glassy states with a completely disordered trimer configuration in the stacking structure. Through PDF analysis using this model, we show that the synthesis method can yield two types of low-temperature stacking structures: a completely disordered stacking structure and a short-range order in the stacking structure. The phase transition temperature of the former sample is about 15 K lower than that of the latter. We discuss that this is due to the strong trimer frustration that appears in the sample without short-range order, which suppresses the phase transition temperature, similar to the frustration effect in conventional spin systems.
\end{abstract}

\maketitle


In transition metal compounds with orbital degrees of freedom, transition metal ions often spontaneously assemble to form ``molecules'' in solids at low temperatures \cite{CuIr2S4, MgTi2O4, LiRh2O4, LiRh2O4-2, AlV2O4, LiMoO2, Li033VS2, CsW2O6, Li2RuO3, LiVS2-1, LiVS2-2, LiVO2-kj, AlV2O4-2}. Since electrons are trapped and localized in bonding orbitals during molecular formation, molecular formation generally appears as a drastic first-order transition to a nonmagnetic insulating state with a large entropy change \cite{LiVS2-1,LiVO2-Tian,LiRh2O4,VO2_entoropy}. Because of these properties, molecular formation has attracted considerable attention not only from the perspective of the fundamental physics of strongly correlated electrons \cite{Khomskii_Mizokawa,Khomskii_Sergey_review} and low dimensionality \cite{Whangbo_Canadell}, but also from an applied perspective, such as sensors \cite{sensor} and phase-change materials \cite{VO2_Muramoto,LiVO2_PCM,Li033VS2,LiVO2-Tian,Niitaka}. 
Molecular formation in solids significantly alters the surrounding crystalline field, supporting the emergence of long-range configurations with the most stable lattice energies at low temperatures. If there are many patterns of molecular arrangement with similar lattice energies and these patterns are not uniquely determined, what structures and physical properties can arise from these ground-state frustrations?

\begin{figure}
\includegraphics[width=85mm]{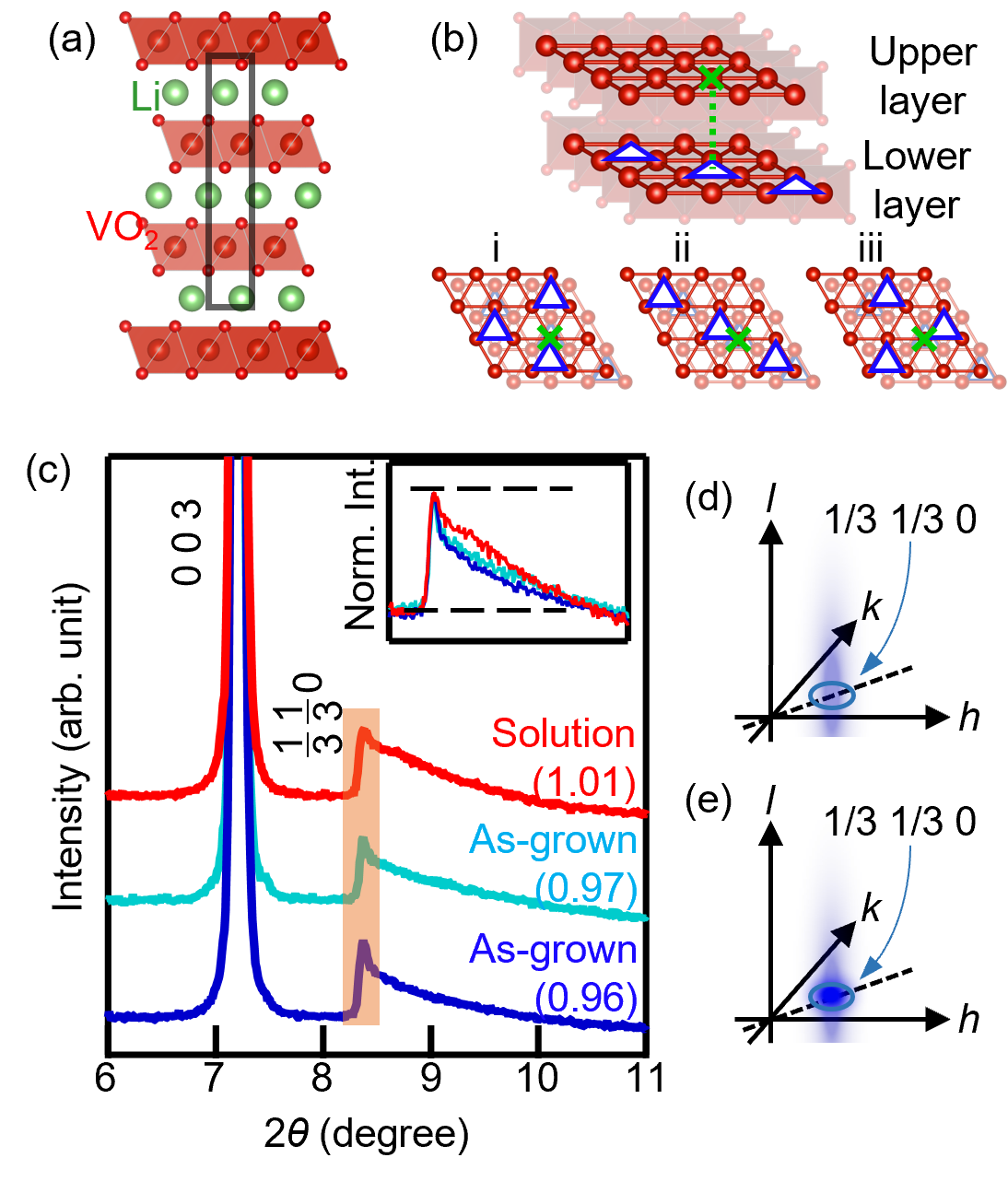}
\caption{\label{fig:Fig1} (a) Horizontal view of the crystal structure of layered LiVO$_2$. (b) Three energetically equivalent trimer patterns in relation to adjacent VO$_2$ layers. (c) Part of X-ray powder diffraction data. The highlighted area corresponds to 1/3 1/3 0. Compared to the solution sample, a sharp peak is clearly generated in this area in the as-grown samples. The inset of the graph shows data normalized so that the background and peak tops (both indicated by dashed lines) are aligned to clarify the shape of the superlattice peaks. (d) Schematic of superlattice reflections of a solution sample with uniform streaks in the $c^*$ direction. (e) Schematic of superlattice reflection of as-grown sample about to condense to 1/3 1/3 0.}
\end{figure}

Layered LiVO$_2$ with a two-dimensional triangular lattice may provide a fascinating playground for studying such an issue \cite{LiVO2-kj}. As shown in Figure~\ref{fig:Fig1}(a), LiVO$_2$ has a stacking structure with three periodic layers, which undergoes a nonmagnetic-paramagnetic transition at about 500 K upon heating \cite{LiVO2-Tian}. The low-temperature nonmagnetism is due to the long-range trimerization of vanadium on in-plane triangular lattice. On the other hand, however, as shown in Figure~\ref{fig:Fig1}(b), there are three degrees of freedom in the arrangement of the trimer for adjacent VO$_2$ layers, and the energies of the lattice structures of these three patterns (i-iii) are equivalent. As a result, the ordered structure should not be uniquely determined. Such a trimer frustration state may seem analogous to spin frustration \cite{Balents, spinglass, SrCu2(BO3)2, CdCr2O4, Tb2Ti2O7, Ba3CuSb2O9, electricalfrust}, but is purely a frustrated state of structural origin. Therefore, it is an exciting research challenge to explore the new electronic phases and physical properties that emerge from ``trimer frustration,'' as in conventional spin systems with geometric frustration. However, this requires techniques for accurately modeling the structural state of LiVO$_2$ and evaluating it with experimental methods.

Here, we report on the modeling of the random trimer arrangement in the stacking direction, and the PDF analysis of LiVO$_2$ based on this model. Our analysis reveals important differences in the local structure of the low-temperature phase of LiVO$_2$ depending on the synthesis method. Samples synthesized by a combination of solid- and solution-reaction methods show completely disordered glassy trimer arrangement, while those synthesized by the solid-phase reaction method show short-range order in the trimer arrangement in the stacking direction. DSC measurements of both samples revealed that the entropy change associated with the phase transition is maximized as the Li/V ratio approaches 1.0. On the other hand, regardless of the Li/V ratio, the trimerization temperature of the sample with a completely disordered trimer arrangement in the stacking direction was decreased by nearly 15 K compared to the short-range ordered sample. We discuss that these results indicate that strong trimer frustration suppresses the low-temperature phase in samples with a completely disordered trimer arrangement in the stacking direction, similar to the suppression of antiferromagnetic order in spin systems with strong magnetic frustration.

We have grown two types of samples, named ``as-grown samples" and ``solution samples," depending on the synthesis method. ``As-grown samples" were synthesized by solid-state reaction. Li$_2$CO$_3$ (99.9\%) and V$_2$O$_3$ (99.99\%) were mixed in the ratio of Li/V $\sim$ 1.00, placed on an alumina boat, and sintered at 625 $\tccentigrade$ for 24 hours with H$_2$/Ar=5\% gas flowing. The obtained samples were regrinded and sintered at 750 $\tccentigrade$ for 12 hours with H$_2$/Ar=5\% gas flowing. The ``solution sample" was obtained by immersing it in a large excess of 0.2 $M$ $n$-BuLi/Hexane solution for 24 hours under an Ar atmosphere in a glove box after the solid phase reaction so that the Li content was almost 1.0. The Li content of these samples was evaluated by ICP measurement. In the following, these samples used in the experiments are labeled ``as-grown(0.97)", ``as-grown(0.96)", ``as-grown(1.01)", and ``solution(1.01)" according to the Li content estimated by ICP measurement.

ICP measurement was performed using a SPECTRO ARCOS MV130 (Hitachi High-Tech). The obtained samples were subjected to DSC measurements using a 204 F1 Phoenix (Netzsch). The temperature rise and fall rates were 10 K/min. Synchrotron X-ray diffraction experiments were performed at BL5S2 of the Aichi SR using a quadruple PILATUS 100 K detector at an $E$ = 20 keV X-ray energy. Diffraction experiments to obtain the pair distribution function (PDF) were carried out at BL04B2 of SPring-8. The experiments were performed using $E$ = 61 keV X-rays, and a combination of four CdTe and three Ge point detectors was used. Rietveld and Le Bail analysis were performed by Rietan-FP \cite{RIETAN}. PDF conversion was performed using a dedicated package \cite{BL04B2}. After corrections, the PDF was obtained by Fourier transform with $0.2<Q~(\mathrm{\AA})<25.5$ and $\mathit{\Delta} Q=0.01~(\mathrm{\AA})$. The simulations of the PDF were performed using PDFgui \cite{PDFGUI}. VESTA was used to draw the crystal structure \cite{VESTA}.


Synchrotron X-ray diffraction results revealed that all the samples used in this study are almost single-phase. Details of the Rietveld analysis are summarized in Supplemental Information \cite{SI}. Figure~\ref{fig:Fig1}(c) shows a part of the powder X-ray diffraction image at 300 K. For the solution(1.01) sample, a sawtooth 1/3 1/3 0 superlattice peak appears. The superlattice peaks are due to the 1/3 1/3 0 superlattice peaks appearing as diffuse streaks in the $l$ direction as shown in Figure~\ref{fig:Fig1}(d), indicating that the trimer is periodically aligned in the plane and randomly aligned in the stacking direction. This has been reported previously \cite{LiVO2-kj}. On the other hand, as shown in Figure~\ref{fig:Fig1}(c), the as-grown samples maintain the sawtooth superlattice peak but generate a sharper peak at 1/3 1/3 0 than the solution(1.01) sample. This suggests that the diffuse streak condenses at 1/3 1/3 0, as shown in Figure~\ref{fig:Fig1}(e), and short-range order develops in the trimer arrangement pattern along the stacking direction in the as-grown sample.

\begin{figure}
\includegraphics[width=85mm]{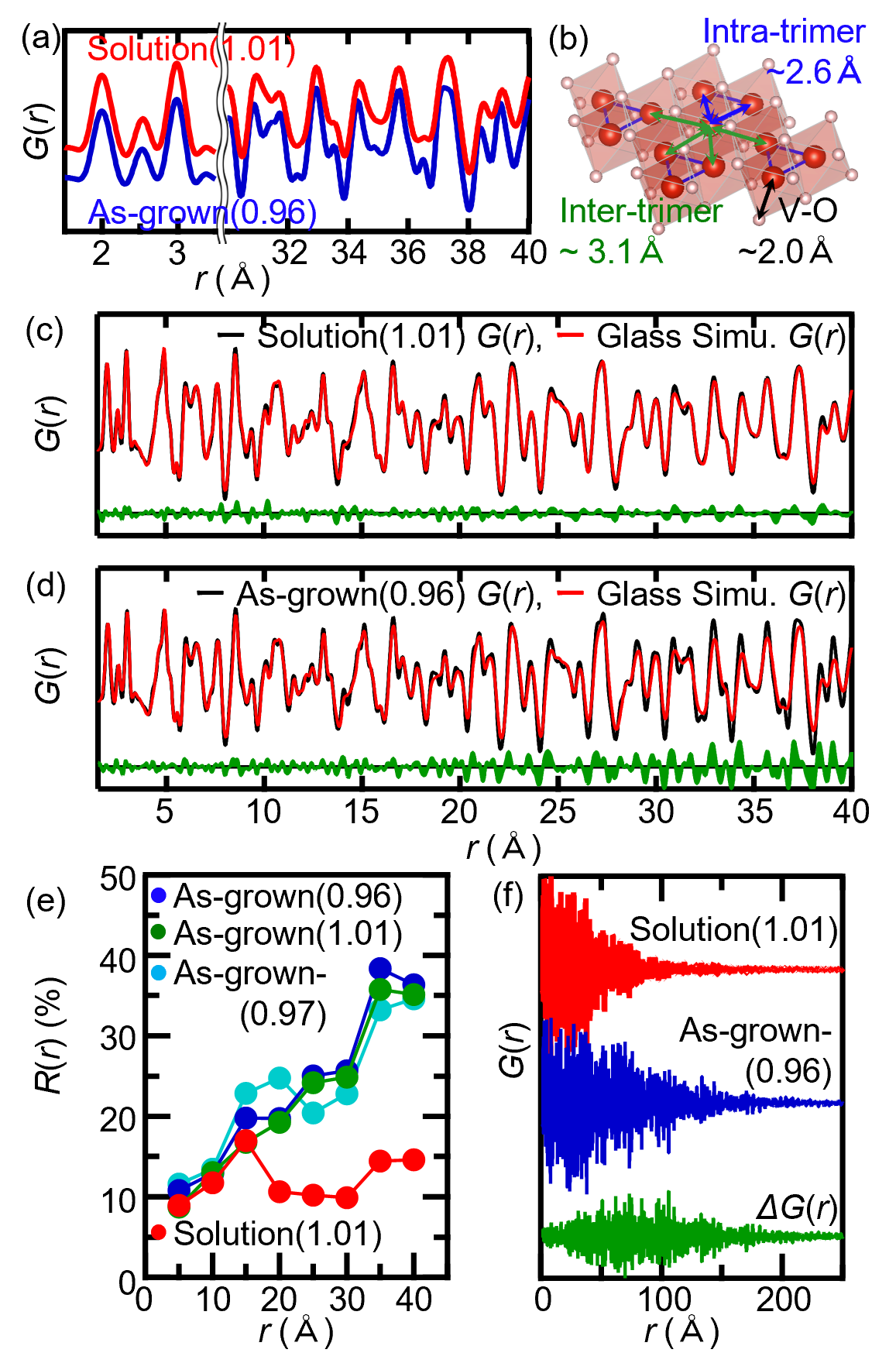}
\caption{\label{fig:Fig2} (a) PDF data at $1.5<r(\mathrm{\AA})<3.5$ and $30<r(\mathrm{\AA})<40$. (b) Distance between atoms in a VO$_2$ plane. (c,d) $G$($r$) pattern of (c) solution(1.01) and (d) as-grown(0.96) fitted by glass simulation data. (e) Confidence $R$($r$) of the refinement obtained in various $r$ regions. (f) $G$($r$) of solution(1.01) and as-grown(0.96) and their difference $\mathit{\Delta}$$G$($r$).}
\end{figure}

The development of short-range order in as-grown samples can be confirmed by pair distribution function (PDF) analysis. Figure~\ref{fig:Fig2}(a) is a magnified view of the PDF data in the low and high $r$ regions. The three peaks occurring in the low $r$ region indicate the nearest-neighbor V-O distance, intra-trimer V-V distance, and inter-trimer V-V distance as shown in Figure~\ref{fig:Fig2}(b). The PDF data of the solution and as-grown samples are similar in this $r$ region, indicating that they have similar in-plane structures. On the other hand, the magnified view of the high $r$ region shows that the PDF patterns of both samples are very different. The PDF spectrum of the solution (1.01) sample shows a ``glassy" pattern of broadening peaks. This reflects the lack of order in the arrangement of the trimer in the stacking direction. On the other hand, in the as-grown sample, the peak shapes are sharper than in the solution sample, the peaks are clearly separated from each other, and even the peaks with weaker intensities are clearly recognizable.


To investigate short-range ordering, PDF simulation patterns in the trimer glassy state were created and fitted to the PDF data. Since three trimer patterns appear in each VO$_2$ layer, 3$^n$ trimer ordered structures appear per $n$ VO$_2$ layers. The glass pattern simulations were created by calculating all of the PDF patterns produced by the 3$^n$ possible patterns and adding them together with a weight of 1/3$^n$ \cite{LiVO2-kj}. As shown in Figure~\ref{fig:Fig2}(c), the simulation data are in good agreement with the experimental PDF data in the range $1.5<r(\mathrm{\AA})<40$, indicating the appearance of a trimer glassy state in solution(1.01). On the other hand, as shown in Figure~\ref{fig:Fig2}(d), for the as-grown samples, a large residual appears above 15 $\mathrm{\AA}$, which corresponds to the thickness of three layers, and the residual expands as $r$ increases. This result can be understood as follows. First, the reduced $G(r)$ at $1.5<r(\mathrm{\AA})<4.0$ contains mainly the component corresponding to the in-plane interatomic distance. The reduced $G(r)$ at $4.0<r(\mathrm{\AA})<15$ contains information on the distances between the atoms in the nearest and next nearest VO$_2$ layers. If there is a short-range order in the trimer arrangement along the stacking direction, large residuals are likely to appear at $4.0<r(\mathrm{\AA})<15$. However, as shown in Figure~\ref{fig:Fig1}(b), the relationships between the trimer in a VO$_2$ layer and the three trimer patterns in the nearest and next-nearest VO$_2$ layers are all equivalent, resulting in the same PDF pattern regardless of which trimer pattern appears. In other words, the PDF pattern at $4.0<r(\mathrm{\AA})<15$ is constant regardless of the presence or absence of short-range order. When the correlation length of the short-range order is longer than 15$~\mathrm{\AA}$, large residuals appear in the region $r(\mathrm{\AA})>15$. This occurs in the fitting of as-grown samples.

In order to quantitatively investigate the agreement between the simulated data and the experimentally obtained PDF data, we defined the following evaluation function $R(r)$,
\begin{equation}
{R(r)=\sqrt{\frac{\sum^r_{r'=r-5}\left\{G(r')_{\mathrm{exp.}}-s\cdot~G(r')_{\mathrm{calc.}}\right\}^2}{\sum^r_{r'=r-5}G(r')_{\mathrm{exp.}}^2}}}\label{eq:one}.
\end{equation}
where $G(r)_{\mathrm{exp.}}$ is the experimental value and $G(r)_{\mathrm{calc.}}$ is the simulated reduced PDF data of the trimer glass pattern. Details on how to obtain $G(r)_{\mathrm{calc.}}$ are described in Supplemental Information \cite{SI}. $s$ is a dimensionless scale factor to normalize $G(r)_{\mathrm{exp.}}$ and $G(r)_{\mathrm{calc.}}$. The value of $s$ that minimizes $R(5)$ was defined as the sample-specific scale factor and used to calculate $R(r)$ in various $r$ ranges. Equation~(\ref{eq:one}) is based on the so-called box-car refinement concept and is useful for estimating the correlation length of the short-range order in as-grown samples. The results are shown in Figure~\ref{fig:Fig2}(e), where the $R($r$)$ values for the solution(1.01) sample remain low in all $r$ regions analyzed, while the $R(r)$ values of the as-grown samples tend to increase uniformly at $r(\mathrm{\AA})\geq15$. In sufficiently large $r(\mathrm{\AA})$ regions, well beyond the correlation length of the short-range order, there is no correlation between distant atoms. Therefore, in sufficiently large $r(\mathrm{\AA})$ regions, the reduced PDF data for the solution and as-grown samples should again agree well. This can be roughly inferred from the $r$ dependence of the difference between the PDF data of the two samples. As shown in Figure~\ref{fig:Fig2}(f), the residuals clearly decrease in the region above 100 $\mathrm{\AA}$, which roughly corresponds to the correlation length.


\begin{figure}
\includegraphics[width=85mm]{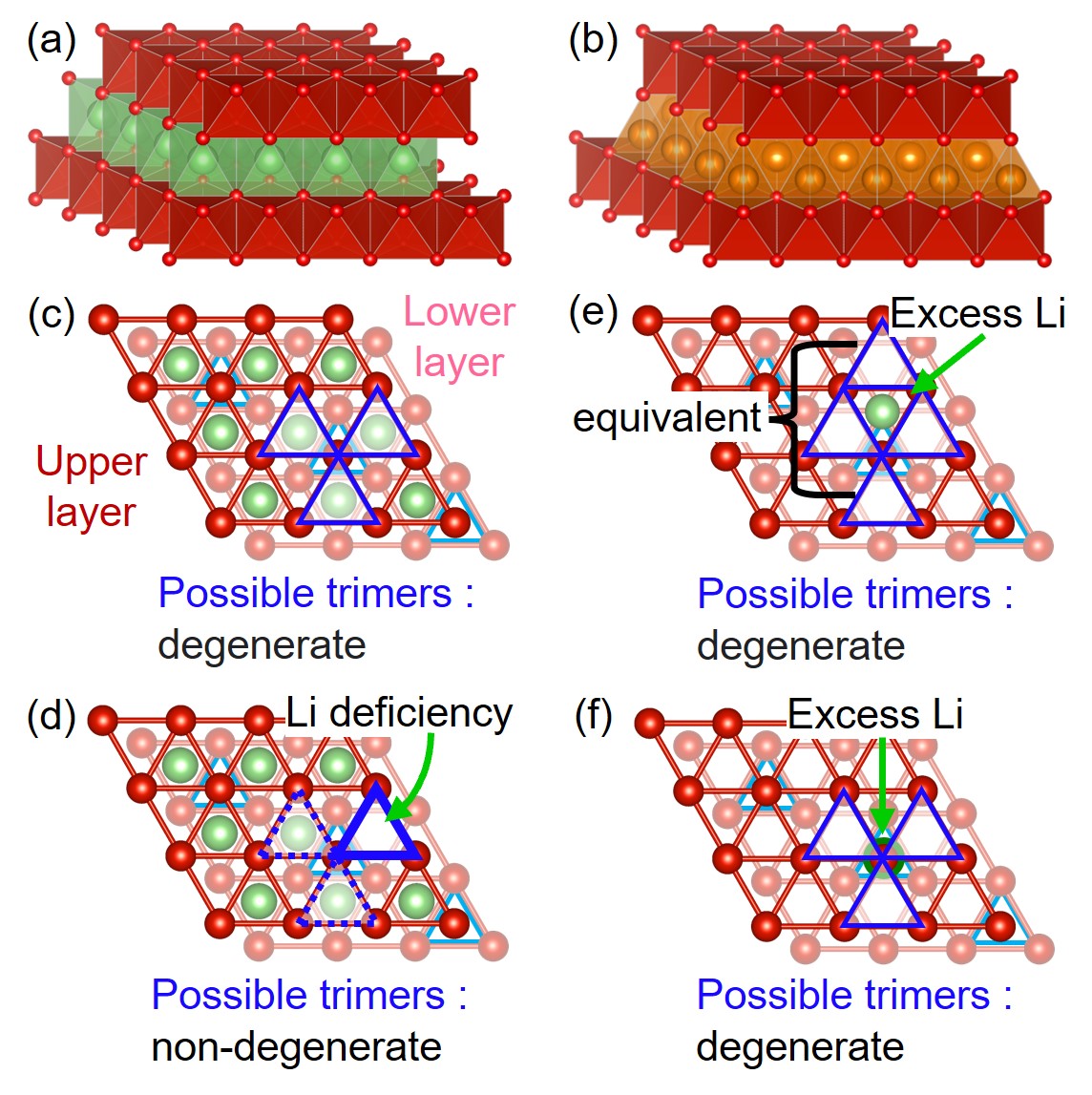}
\caption{\label{fig:Fig4} (a) Octahedral site with majority Li ions. (b) Tetrahedral site with excess Li ions. (c) Relationship between Li ion and three possible trimer configuration. Li ions are located between the upper and lower VO$_2$ layers. (d) Relationship to the trimer when Li is deficient. (e,f) Relation between Li ions at the tetrahedral sites and (e) upper and (f) lower VO$_2$ layers.}
\end{figure}

The cause of the difference between the solution and as-grown samples is not clear but is speculated as follows. In the as-grown sample, the lattice energy degeneracy due to the numerous stacking patterns is thought to be lifted, and short-range order is realized in the trimer arrangement in the stacking direction. Since there is no significant difference in the basic lattice structure between the solution and as-grown samples, the interlayer Li ion sites are likely responsible. Since the short-range order also appears in as-grown(1.01), the lack of Li ions is not the origin of the short-range order. A possible explanation is the disorder of the Li ion sites. As shown in Figure~\ref{fig:Fig4}(a), almost all Li ions are ordered into octahedral sites between VO$_2$ layers, but there are also tetrahedral sites where extra Li ions can enter as shown in Figure~\ref{fig:Fig4}(b). If the Li ions fully occupy the octahedral site, the Coulomb potential of the Li ions on the VO$_2$ layer is uniform and the lattice energy degeneracy associated with trimerization is preserved, so strong frustration is expected, as shown in Figure~\ref{fig:Fig4}(c). However, if deficiencies or other disturbances exist at the octahedral site, the random potential should lift the lattice energy degeneracy and produce a stable ordered structure, as shown in Figure~\ref{fig:Fig4}(d). Interestingly, the random insertion of Li ions into the tetrahedral site does not lift the lattice energy degeneracy because it gives equal random potentials for the three trimer patterns that appear in the neighboring VO$_2$ layers as shown in Figures~\ref{fig:Fig4}(e) and (f). Therefore, we speculate that the solution reaction with $n$-BuLi may have an annealing effect that encourages Li ions to move between the layers to fully occupy the low-potential octahedral sites, in addition to the effect of adjusting the Li ion content. How does the presence or absence of such short-range order affect electronic properties?

\begin{figure}
\includegraphics[width=85mm]{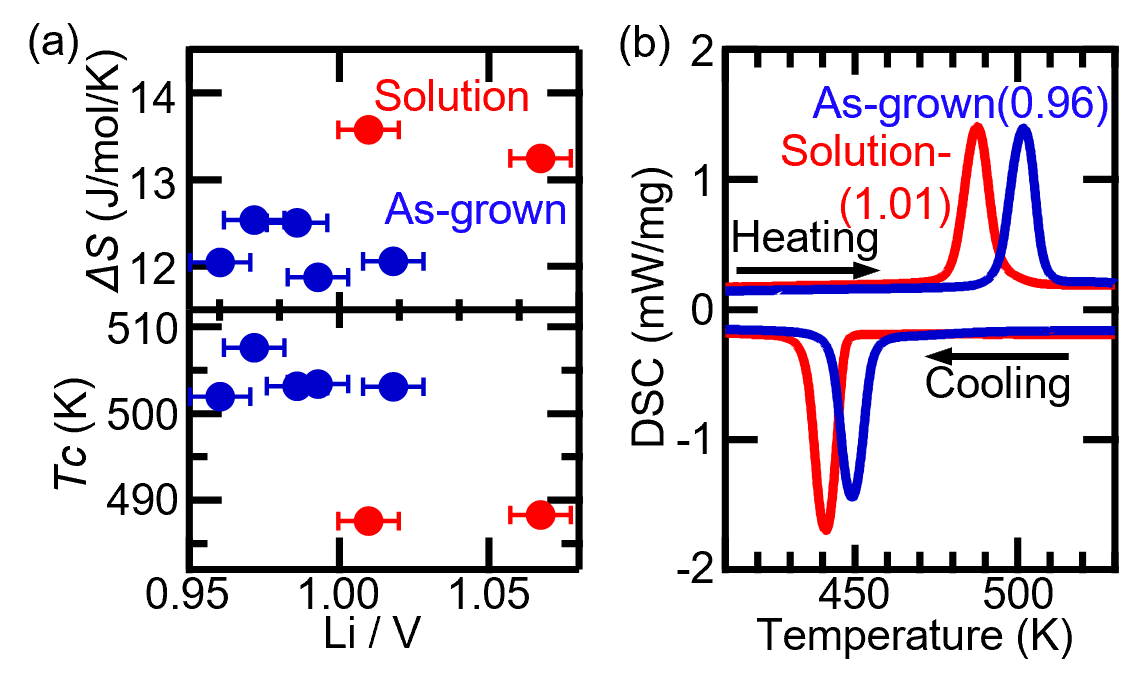}
\caption{\label{fig:Fig3} (a) (upper) Entropy change measured by DSC on heating process, and (lower) the phase transition temperatures. The Li/V ratio is determined by ICP measurements. (b) DSC data for as-grown and solution samples.}
\end{figure}

To explore the effect of such short-range ordering on the physical properties, DSC measurements were performed, and as shown in Figures~\ref{fig:Fig3}(a) and (b), the entropy change associated with the phase transition was always larger in the solution sample than in the as-grown sample. This is understood to be due to the optimization of the electronic state of V in the solution sample as a result of controlling the Li content with $n$-BuLi. On the other hand, in contrast to the trend of the entropy change, the phase transition temperature of the solution sample is $\sim$ 15 K lower than that of the as-grown sample. This result is clearly inconsistent with the entropy change data suggesting stabilization of the trimer structure.

These results seem to indicate that the phase transition temperature is suppressed in samples without the short-range ordering of the trimer arrangement in the stacking direction compared to samples with short-range ordering. This is reminiscent of frustration effects in spin systems. The present trimer frustration state in LiVO$_2$ is a unique state formed by the coupling of electrons and lattice degrees of freedom, and is a consequence of pure lattice ordering. Nevertheless, it is similar to spin systems in that the presence of frustration suppresses the phase transition, and weakening the frustration induces an ordered state and increases the transition temperature. This seems to indicate that frustration effects similar to those in spin systems can be realized in lattice systems.


In spin-frustrated systems, the strength of frustration is quantified by the absolute value of the ratio of the Weiss temperature to the N\'{e}el temperature (frustration factor). In the present trimer frustration, the spin gap estimated from NMR measurements may be an indicator of the strength of the frustration. From previous NMR measurements on LiVO$_2$, the spin gap in the low-temperature phase is estimated to be $\mathit{\Delta}$ $\sim$ 3400 K \cite{LiVO2_NMR_Kawasaki} (1600 K \cite{LiVO2_NMR_Onoda}), which is much larger than the phase transition temperature of LiVO$_2$, $T_c$ $\sim$ 500 K. This seems to indicate that the trimer transition temperature in LiVO$_2$ is strongly suppressed.

One might attribute the energy differences to the presence of local orbital degeneracy lifted state (ODL) that develops prior at high temperatures, as recent PDF studies of the local structure have found in many systems that form orbital molecules at low temperatures \cite{LiVS2-2, Li2RuO3_2, CuIr2S4_2, MgTi2O4_2, MgTi2O4_3, NaTiSi2O6, Attfield, AlV2O4-2}. This may be the case for LiVS$_2$, an analog of LiVO$_2$. LiVS$_2$ has a different stacking structure than LiVO$_2$ and no trimer frustration \cite{LiVO2-kj}, but a trimer transition occurs at 314 K with a gap of $\mathit{\Delta}$ $\sim$ 1900 K \cite{LiVS2-1,LiVS2_NMR_Tanaka2,LiVS2_NMR_Tanaka1}. Above the phase transition temperature, PDF analysis shows that a zigzag chain-like short-range order develops, suggesting that orbital degeneracy is already locally lifted at high temperatures \cite{LiVS2-2}. However, this is not the case for LiVO$_2$. Our previous PDF studies on LiVO$_2$ have shown that no such short-range order develops above the phase transition temperature in LiVO$_2$ \cite{LiVO2-kj}. The above indicates that LiVO$_2$ and LiVS$_2$ are not similar and each has unique physics for trimer formation.

It should be noted that we were able to address such a physics of trimer frustration because of our success in modeling the trimer glassy state of LiVO$_2$ and identifying its structure by PDF analysis. The existence of vanadium trimer formation in LiVO$_2$ was pointed out more than half a century ago based on the lattice symmetry of the low-temperature phase \cite{LiVO2_Goodenough}. Subsequent studies have confirmed the V-V distance splitting associated with trimer formation by EXAFS \cite{Imai} and PDF analysis \cite{Pourpoint}, electron diffraction analysis \cite{LiVO2-Tian}, and the NMR measurement using a single crystalline sample \cite{Jinno}. All of these results support the in-plane appearance of the trimer, but the identification of the crystal structure containing the trimer had not been successful for more than half a century. This is because the trimer disorder in the stacking direction, intrinsic to LiVO$_2$, has not been properly modeled. Coupled with glassy state modeling, PDF analysis was essential in the present results to reveal that LiVO$_2$ is the playground where the new physics of trimer frustration emerges. This achievement can never be revealed by conventional average structure analysis, and may point to a new direction in structural analysis.


Finally, we point out the importance of this trimer frustration in terms of applications. The latent heat calculated from the entropy change of LiVO$_2$ ($\mathit{\Delta}H$ $\sim$ 326 Jcc$^{-1}$) is equivalent to that of H$_2$O ($\mathit{\Delta}H$ $\sim$ 306 Jcc$^{-1}$) and is promising as a phase change material (PCM) product \cite{VO2_Muramoto,LiVO2_PCM,Li033VS2,LiVO2-Tian,Niitaka}. If the phase transition temperature can be manipulated by controlling trimer frustration, it could be a PCM material that can be used at various temperatures. Such studies are beyond the scope of this study, but they clearly demonstrate the importance of both the fundamental and applied aspects of trimer frustration.


\begin{acknowledgments}
All authors thank Dr. K. Ohara and S. Hashimoto for fruitful discussion. The work leading to these results has received funding from the Grant in Aid for Scientific Research (Nos.~JP17K17793, JP20H02604, JP21K18599, JP21J21236
). This work was carried out under the Visiting Researcher’s Program of the Institute for Solid State Physics, the University of Tokyo, and the Collaborative Research Projects of Laboratory for Materials and Structures, Institute of Innovative Research, Tokyo Institute of Technology. PXRD experiments were conducted at the BL5S2 of Aichi Synchrotron Radiation Center, Aichi Science and Technology Foundation, Aichi, Japan (Proposals No. 202002076, No. 202104111, No. 2021L3002 and No. 202105170), and at the BL04B2 of SPring-8, Hyogo, Japan (Proposals No. 2018B1128, No. 2019A1218, No. 2021A1112 and No. 2021B1119).
\end{acknowledgments}

\appendix

\nocite{*}

\bibliography{references}

\begin{thebibliography}{48}
\bibitem [{SI()}]{SI}%
  \BibitemOpen
  \bibinfo {title} {See {S}upplemental {M}aterial at [url will be inserted by
  publisher for experimental details and structural aspects.}\BibitemShut
  {Stop}%
\end{thebibliography}

\end{document}